\documentclass[10pt,twocolumn,twoside]{IEEEtran}
\usepackage{graphicx}
\usepackage{caption}
\usepackage{subfigure}
\usepackage{amsmath}
\usepackage{amssymb}
\usepackage{bm}
\usepackage{color}
\usepackage{url}
\usepackage{epsfig,latexsym}
\usepackage{float}
\usepackage{indentfirst}
\usepackage{amsmath}
\usepackage{amssymb}
\usepackage{times}
\usepackage{subfigure}
\usepackage{cite}
\usepackage{amsfonts,amssymb,amsmath}
\usepackage{subeqnarray}
\usepackage{color}
\usepackage{color}
\definecolor{green}{rgb}{0.796,0.948,0.816}
\usepackage{epstopdf}

\makeatletter

\newcommand{\Rmnum}[1]{\expandafter\@slowromancap\romannumeral #1@}
\makeatother
\usepackage{psfrag}
\usepackage{xcolor}
\usepackage{srcltx}
\usepackage{enumerate}
\usepackage{extarrows}

\graphicspath{{./fig/}}

\begin{document}

\title{{\huge Physical Layer Security Jamming: Theoretical Limits and Practical Designs in Wireless Networks}}
 \markboth{\textit{A Manuscript accepted in The IEEE  Access} }{}
\author{Kanapathippillai Cumanan, Hong Xing, Peng Xu, Gan Zheng, Xuchu Dai, Arumugam Nallanathan, Zhiguo Ding and George K. Karagiannidis
\thanks{
K. Cumanan is with the Department of Electronics, University of York, YO10 5DD, UK.

Z. Ding is with the School of Computing and Communications, Lancaster University Lancaster, LA1 4WA, UK.

H. Xing and A. Nallanathan are with the Department of Informatics, King's College, London, UK.

P. Xu and X. Dai are with the Department of Electronic Engineering and
Information Science, University of Science and Technology of China, China.

G. Zheng is with the Wolfson School of Mechanical, Electrical and Manufacturing Engineering, Loughborough University, UK.

G. K. Karagiannidis is with the Department of Electrical \& Computer Engineering, Aristotle University of Thessaloniki, Thessaloniki, Greece.
}
}
\date{}
\maketitle
\begin{abstract}
Physical layer security has been recently recognized as a promising new design paradigm to provide security in wireless networks. In addition to the existing conventional cryptographic methods, physical layer security exploits the dynamics of fading channels to enhance secured wireless links. In this approach, jamming plays a key role by generating noise signals to confuse the potential eavesdroppers, and significantly improves quality and reliability of secure communications between legitimate terminals. This article presents theoretical limits and practical designs of jamming approaches for physical layer security. In particular, the theoretical limits explore the achievable secrecy rates of user cooperation based jamming whilst the centralized, and game theoretic based precoding techniques are reviewed for practical implementations. In addition, the emerging wireless energy harvesting techniques are exploited to harvest the required energy to transmit jamming signals. Future directions of these approaches, and the associated research challenges are also briefly outlined.
\end{abstract}
\section{Introduction}
\indent In wireless communications, the exponential growth of mobile traffic and newly emerging wireless applications introduce different security risks due to their broadcasting nature. The secured communication links in traditional wireless networks are established through conventional cryptographic methods. However, these methods impose different challenges in terms of key exchange and distribution, especially in the current trend of dynamic network configurations. Recently, physical layer security has been recognised as one of the potential solutions to enhance security in wireless networks by exploiting characteristics of wireless channels \cite{MISOME,HH_Chen_WC_J11,Cuma_TVT_J14,Qaraqe_Comm_lett_J16}. In addition, this novel paradigm complements the conventional cryptographic methods, and well suits for dynamic networks and distributed processing techniques.\\
\indent Physical layer security jamming is a well known approach to enhance the quality of secure wireless transmissions \cite{tekin2008general,mukherjee2014survey,Zheng_TSP,Xia_TWC_J15,Xia_Comm_Mag_J16}. In this technique, additional jamming signals are transmitted to confuse the potential eavesdroppers or to degrade the decoding capability of the unintended receivers. These jamming signals can be introduced by embedding them with the intended signals, which are referred as artificial noise (AN) approach in the literature. On the other hand, a receiver can be also used to transmit jamming signals with the help of full duplex (FD) radios, which have the capability to simultaneously transmit and receive the signals. Hence, FD receiver can be exploited to receive the required signals while sending jamming signals at the same time to confuse the eavesdroppers \cite{Zheng_TSP}. However, this transmit and receive jamming scheme might not be possible under all circumstances due to limited available number of antennas and the strong self-interference (SI). In this scenario, the external nodes can be employed to send jamming signals, where they could be relay nodes or private jammers \cite{Han_TVT_J12,Chu_TVT_J14,Cuma_JSTSP_J16}. In case of private jammers, they could introduce charges for their dedicated jamming services. The problems associated with these private jammers can be formulated into different game theoretic problems by considering the legitimate nodes and the private jammers as the players of the game.\\
\indent This article focuses on physical layer security jamming techniques based on user cooperations and external nodes. Firstly, theoretical limits of jamming through user cooperation is presented, and then multi-antenna based jamming techniques are reviewed by exploiting their spatial diversity and degrees of freedom (DoF). For example, the advantages of jamming with multi-antenna transmitter can be easily demonstrated by appropriately designing beamformers such that it would cause a significant interference to the eavesdroppers while no or less interference leakage to the intended receivers. However, the study of theoretical limits of jamming and the practical designs are necessary to achieve the optimal performance in secrecy networks. This article presents these theoretical limits and design approaches as follows. First, theoretical limits of user cooperation based jamming are explored. Then, centralized and game theoretic based multiple-input multiple-output (MIMO) transmit and receive precoding techniques are discussed to provide efficient jamming services. In addition, wireless energy harvesting (WEH) based jamming techniques are presented through the recent advancement in simultaneous wireless information and power transfer (SWIPT) concept. Finally, future research challenges of jamming schemes are briefly discussed.
\section{Information-theoretic Limits of Jamming}
\begin{figure}[t]
\begin{center}
   \subfigure[]{\includegraphics[width=0.47\textwidth]{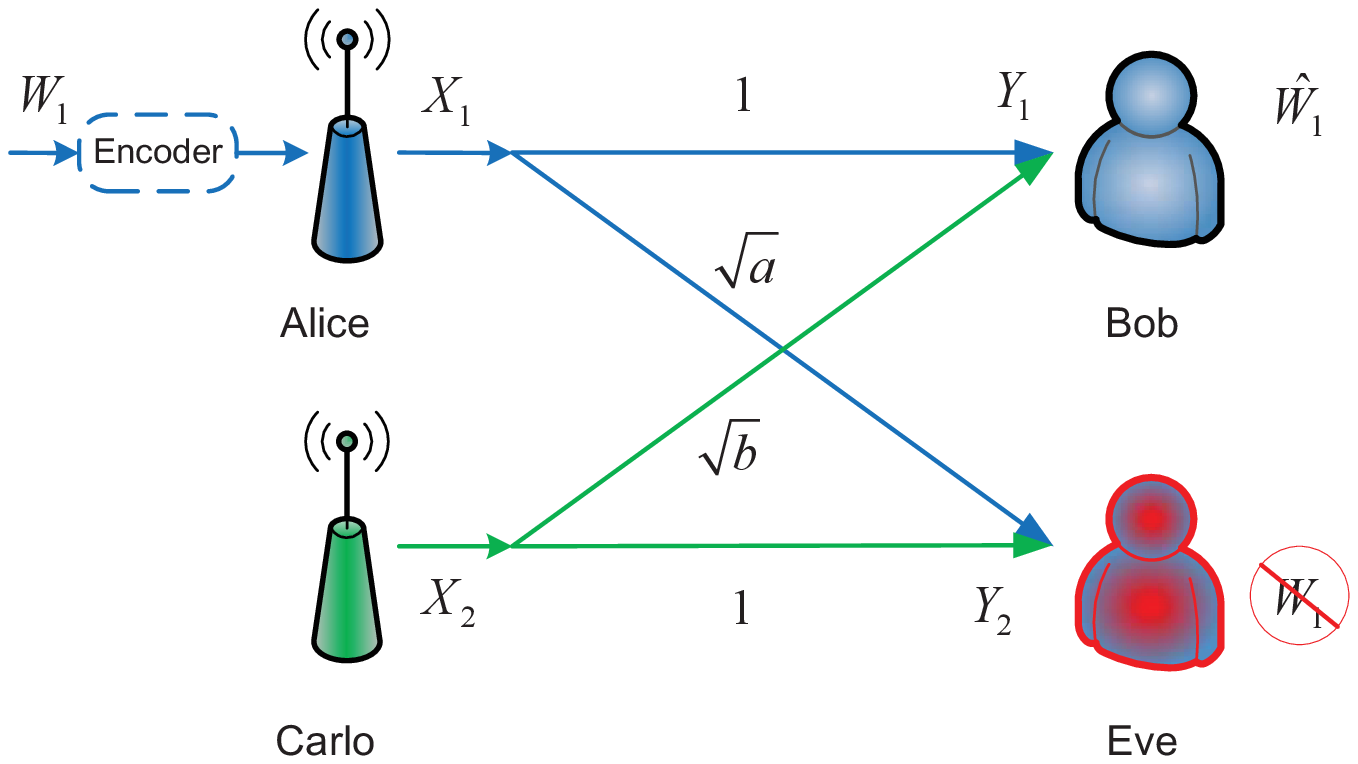}}
   \subfigure[]{\includegraphics[width=0.47\textwidth]{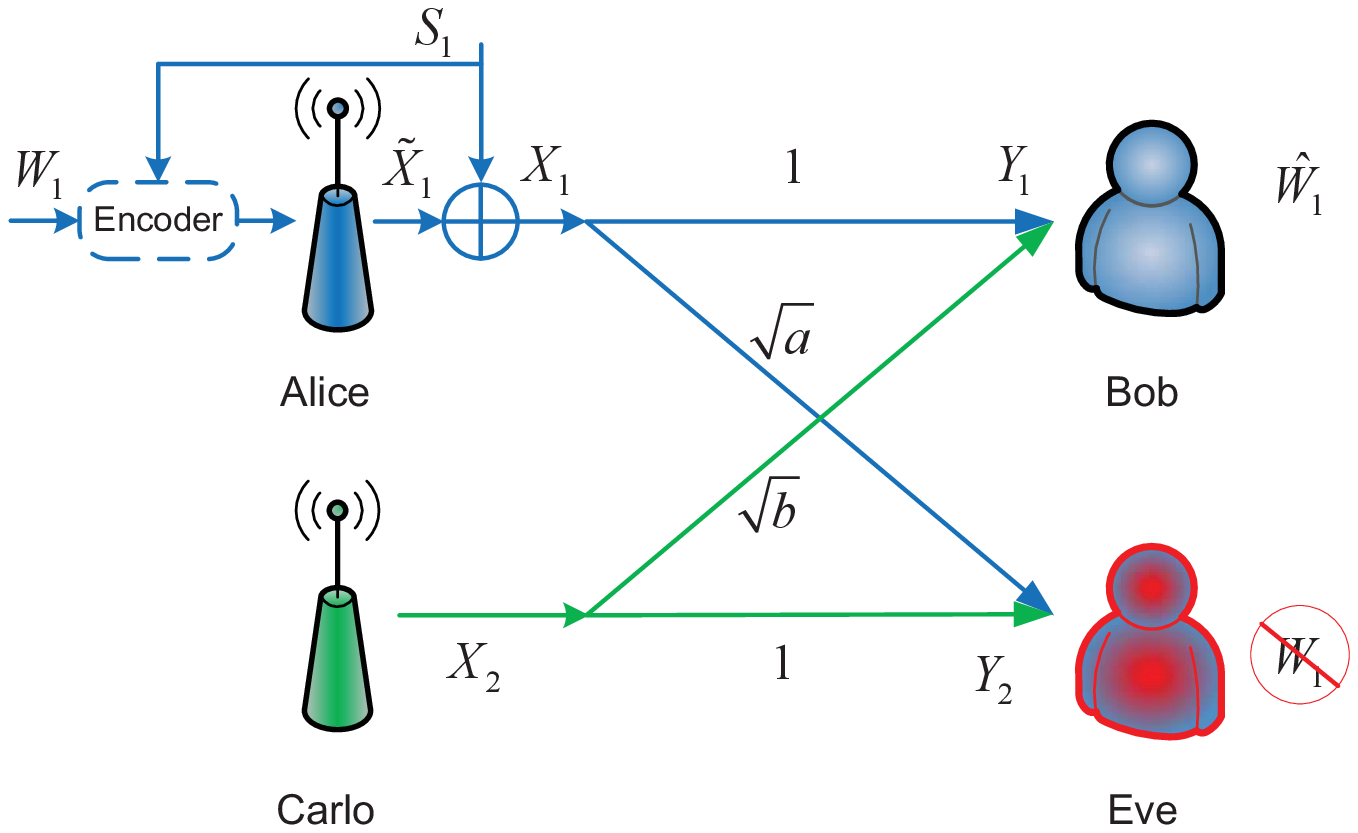}}
\end{center}
\caption{(a) The  wiretap channel with a helping interferer (WT-HI).
(b) The  WT-HI channel
with artificial state.}\label{figure1}
\end{figure}
\indent In this section, we present the theoretical limits of cooperative-users based jamming for physical layer security. The concept of cooperative jamming (CJ) was introduced in \cite{tekin2008general} for a Gaussian multiple
access wiretap channel (GMAC-WT) from an information-theoretic aspect.
In the GMAC-WT channel, multiple legitimate users wish to send secret messages to
a common receiver in the presence of a passive eavesdropper. In order to maximize all the sum secrecy rate,
a user should send pure Gaussian noise as long as the eavesdropping channel from it to the eavesdropper is
stronger than the legitimate channel from it to the intended receiver. The CJ scheme can be illustrated
via a simple two-user GMAC-WT channel as shown in Fig. \ref{figure1}, where a transmitter (Alice)
wishes to send  a secret message
$W_1$ to the intended receiver (Bob) under the help of an interferer  (Carlo), without leaking any information to the eavesdropper (Eve). The channel gains from Alice to Bob and Eve are normalized to  be 1 and $\sqrt{a}$;
the channel gains from Carlo to Bob and Eve are normalized to  be  $\sqrt{b}$ and 1.
Note that the channel gains from Alice to Bob and from Carlo to Eve are different
with each other although both of them are normalized to  1, which is reflected by the fact that both the channels have
different effects on the received signal-noise-ratios  (SNRs)
at  Bob and Eve.
This channel model is also called the wiretap channel with a helping interferer (WT-HI) in \cite{tang2011interference}.
When $b<1$, Carlo can
help Alice and Bob to enhance the security level by sending Gaussian noise that is independent of the
message-carrying signal. This can be interpreted by the fact that Carlo's jamming signal harms Eve more than Bob
when $b<1$, which may improve  the achievable secrecy rate.

 CJ is useful only when $b<1$ in the WT-HI channel in Fig. \ref{figure1} (a). But, when $b>1$, i.e., Carlo has a
 stronger channel to Bob than to Eve, CJ in \cite{tekin2008general} might not be useful.
 In this case, Carlo can still help Alice and Bob using the noise forwarding (NF) scheme in \cite{lai2008relay}.
 In NF, Carlo can randomly choose a codeword from a known codebook with
 an appropriate coding rate such that the confusion signal can be decoded by Bob before
 the decoding of $W_1$, and hence, without affecting the message-carrying signal, still jams the eavesdropping channel.
 An interpretation of  NF  is that the independent confusion codewords
  can bring additional randomness to the channel to enhance the security level.
 The main difference between NF and CJ is that, the former designs the confusion signals with structure that does not jam Bob,
 whereas the latter uses pure noise  that jams Bob and Eve simultaneously.

 Both CJ and NF can be generalized into a unified framework based on the adaptive adjustment of the coding rate at Carlo,
 which is called interference assisted (IA) scheme in \cite{tang2011interference} for the WT-HI channel.
  The main difference between the IA and NF schemes
  is that,
 in the former, Carlo treats its coding rate as a variable, and adaptively adjusts this coding rate to
 maximize the achievable secrecy rate; while in the latter, Carlo always chooses a coding rate such that the interference transmitted by him
 does not affect Bob. Both CJ and NF can be viewed as the special cases of the IA scheme. When the coding rate at Carlo
 is lower than a certain rate such that Bob can decode the interference before decoding the secret message, the IA scheme reduces
 to NF; when the coding rate at Carlo is sufficiently large such that both Bob and Eve have no choice but to treat the
 interference as pure noise, the IA scheme reduces to CJ.

\begin{figure}[t]\centering
    \epsfig{file=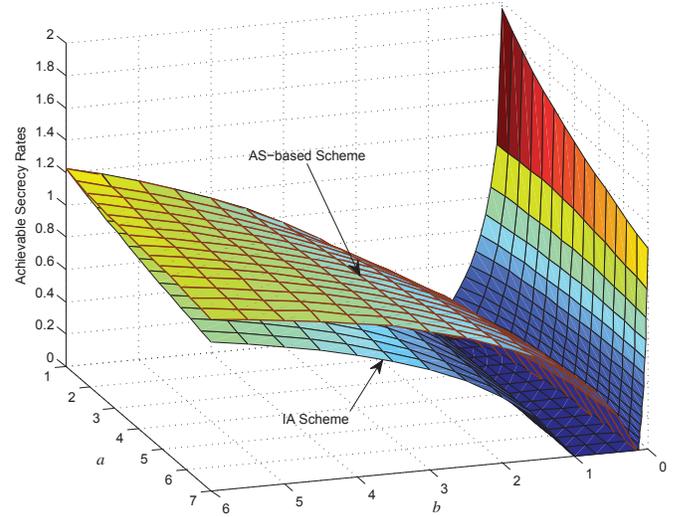, width=0.5\textwidth,clip=}
    \caption{Achievable secrecy rates of the IA and AS-based schemes for the WT-HI channel shown in Fig. \ref{figure1} (a),
    where the maximum transmit power at Alice and Carlo are 20 and 50, respectively. Since
    the IA scheme is a generalization of the CJ and NF scheme, its achievable secrecy rate  is the same as that of
     the CJ scheme when $b<1$ and the same as that of the NF scheme when $b>1$.
     }\label{secrecy_rate}
\end{figure}

 CJ and NF can also be generalized by the channel prefixing scheme in secure communications. As shown in the Gaussian
 WT-HI channel in Fig. \ref{figure1} (a), the confusion signal transmitted by Carlo can consist of two parts, i.e., $X_2=U_2+Z_2$,
 where one part $(U_2)$ is the codeword with the structure following the NF scheme, and
 the other part $(Z_2)$ is pure Gaussian noise following the CJ scheme. In this
 channel prefixing scheme, the channel input $X_2$ at Carlo becomes a noisy version of the interference codeword $U_2$
 in terms of the Markov
 chain $U_2-X_2-Y_1,Y_2$. In addition to Carlo's channel prefixing,
 the work in \cite{xu2014general} also adopted a channel prefixing scheme at Alice based on artificial state (AS), and  the achievable secrecy rate can be further enhanced.
Specifically,  Alice spends part of its transmit power to generate an AS $S_1$,
 i.e., it sets the channel input as  $X_1=\tilde{X}_1+S_1$ as shown in Fig. \ref{figure1} (b).
  Now, the WT-HI channel in Fig. \ref{figure1} (a) becomes
  a WT-HI channel with state information as shown in Fig. \ref{figure1} (b),
    where the random state $S_1$ is known by Alice a priori and  $\tilde{X}_1$ can be viewed as the virtual channel input.
 Then following the concept of dirty paper coding (DPC), we introduce an auxiliary variable
  $U_1=\tilde{X}_1+\beta S_1$ as the
 message-carrying codeword. According to DPC, $S_1$ will not affect the decoding of the message-carrying codeword $U_1$
 at Bob by appropriately setting the value of $\beta$.

Fig. \ref{secrecy_rate} shows
 achievable secrecy rates of the IA and AS-based schemes for the WT-HI channel shown in Fig. \ref{figure1} (a),
    where the maximum normalized transmit power at Alice and Carlo are 20 and 50.
    For both schemes, Gaussian codebooks are adopted, where the message-bared signal $X_1\sim\mathcal{N}(0,P_1)$,
     and the Gaussian interference $X_2\sim\mathcal{N}(0,P_2)$,
     $0\leq P_1\leq 20$, $0\leq P_2\leq 50$. Furthermore, the AS-based scheme sets
    $\tilde{X}_1\sim\mathcal(0,(1-\lambda)P_1)$,  $S_1\sim\mathcal(0,\lambda P_1)$.
    The optimal values of these parameters $(P_1,P_2,\lambda)$ are established via exhaustive searches.
As shown in Fig. \ref{secrecy_rate}, when $a>1$,
the AS at Alice is important to get a larger
secrecy rate if compared with the NF (or IA) scheme.
Particularly, when $b=1$, Carlo can still assist Alice
to achieve a positive secrecy rate, whereas the CJ, NF and IA schemes fail. This can be interpreted by the fact that
 the AS at Alice can associate with the helping interference at Carlo  to further
confuse Eve without affecting Bob. Fig. \ref{secrecy_rate} also shows that the gap between the AS-based and
the NF schemes can be enlarged by increasing $a$ from 1 to 7. Moreover, the AS-based scheme reduces to the CJ (or IA) scheme when
$b<1/a$, thanks to the channel prefixing at Carlo.

The works in \cite{tekin2008general,lai2008relay,tang2011interference,xu2014general} used information-theoretical
Gaussian codebooks with infinite alphabets, which is hard to implement in practice. Alternatively, the work in \cite{he2014structured}
has proposed an achievable scheme based on layered nested lattice codes with an finite alphabets. Interestingly,
unlike traditional communications without secrecy constraints, the use of more practical lattice codes
can outperform the Gaussian codebook for  the channel model in Fig. \ref{figure1} (a) for certain cases.
 Specifically, the scheme in
\cite{he2014structured} can achieve non-zero
 secure degree of freedom (s.d.o.f.) for all values
 of channel gain pair $(\sqrt{a},\sqrt{b})$ except when $ab=1$, whereas the works in  \cite{tekin2008general,lai2008relay,tang2011interference,xu2014general}
 fail to achieve non-zero s.d.o.f.. This is because,
 based on nested lattice codes, the signals transmitted by both the source and
the cooperative jammer can be aligned at the eavesdropper
but remain separable at the intended receiver; whereas
the Gaussian noise transmitted by the cooperative jammer
simultaneously  interferes with the eavesdropper, and
hurts the intended receive.

\section{MIMO Jamming: Centralized Approach}
In this section, we review key results of using MIMO to transmit
judiciously jamming signals with an aim to achieve higher secrecy
rate. Specifically, centralized transmit and receive jamming techniques are discussed in the following subsections.

\subsection{Transmit Jamming}
Consider a basic three-node system, which consists of a transmitter,
an intended receiver and an eavesdropper.  The transmitter has
multiple antennas, while the receiver and the eavesdropper may have
multiple antennas or a single antenna. The secrecy capacity is well
known \cite{MISOME} when the channel state information (CSI) is available to
all nodes. While it is reasonable to assume to have the receiver's
CSI, it is usually unrealistic to obtain  the eavesdropper's CSI. In
the case where only the receiver's CSI is known but the
eavesdropper's CSI is absent, the transmit jamming or AN is an effective means to improve the secrecy rate. As shown in
Fig. \ref{fig:MIMO:jamming} (a), the total transmit signal is split
into two parts. The first one is the information-bearing message
towards the direction of the receiver, which can be designed based on
the receiver's CSI. The second part is the transmit jamming, which is
isotropic Gaussian noise in the orthogonal space of the receiver's
channel. By doing so, the transmit jamming does not affect the
intended receiver but only degrades the quality of the signal
received by the eavesdropper. A remarkable result about this simple
transmit jamming scheme shown in \cite{MISOME} is that, it can
achieve secrecy rate close to the  capacity in the high SNR regime
when the receiver has a single antenna.

The use of transmit jamming consumes some transmit power so it
reduces the SNR at the receiver.
Therefore it is important to allocate power to the
information-bearing signals and the jamming signals properly. While
it is in general difficult to achieve the optimal power allocation,
a general rule of thumb is to allocate more power to jamming when
the receiver's CSI is more accurate or  the number of the
eavesdropper's antennas increases to achieve effective jamming.

\subsection{Receive Jamming}
Transmit jamming is useful but requires the support of multiple
antennas. When the transmitter has a single antenna, is it still
possible to  take advantage of jamming?
Fortunately, there is a positive answer to this question.
One possible solution is to use  jamming at the receiver side to confuse
the eavesdropper. There are different implementations. For
instance, the transmitter repeats the transmission for a certain
times, and the receiver randomly jams the transmissions \cite{iJam}.
Because the eavesdropper does not know which transmission is left
unjammed, it cannot decode the message correctly. One drawback of this
scheme is that, it requires retransmission, which is less bandwidth efficient and may be critical for delay-sensitive applications.
Next we introduce a receive jamming scheme without using retransmissions.

\begin{figure*}[t]
\centering
\includegraphics[width=0.8\textwidth]{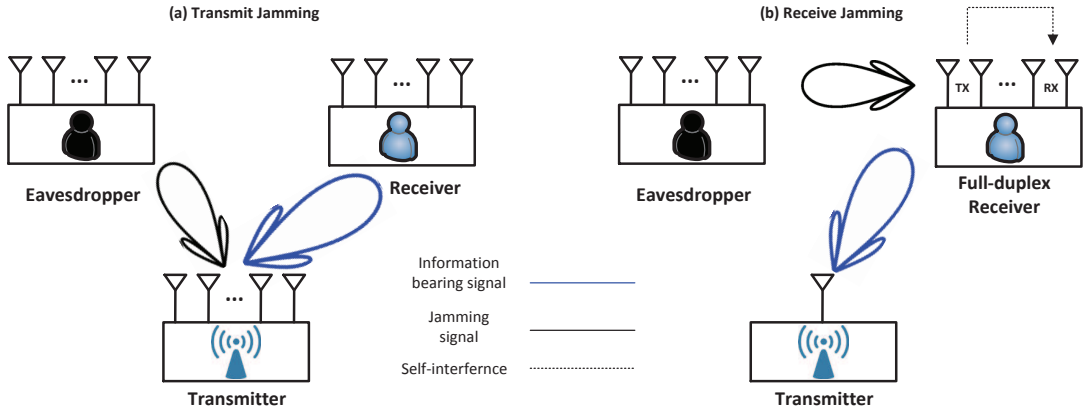}
\caption{MIMO Jamming. (a) Transmit jamming when the transmitter has
multiple antennas; (b) Receive jamming when the receiver has
multiple antennas and works in the FD
mode.}\label{fig:MIMO:jamming}
\end{figure*}

Traditionally a wireless node works in the half-duplex (HD) mode.
The FD operation, which allows a wireless node to
transmit and receive simultaneously in the same frequency band has emerged as an
attractive solution to improve the spectral efficiency. The FD
operation has also shown great potential to improve physical layer
security. As depicted in Fig. \ref{fig:MIMO:jamming}(b), the main
idea is that the intended receiver sends jamming signals to degrade
the eavesdropper's channel and protect its own reception
\cite{Zheng_TSP}. Obviously the receive jamming will affect both the
intended receiver itself and the eavesdropper because of the
resulting SI, but if the SI can be well
controlled/optimized, it will favor the intended receiver. When the
 receiver has  multiple transmit or receive antennas, it can employ
joint transmit and receive beamforming  for simultaneous signal
detection, SI suppression and jamming emission. An interesting
result in  \cite{Zheng_TSP} is   that when the global CSI is
available, the secrecy rate increases unbounded as the SNR goes up,
which is in contrast to the traditional HD case without receive
jamming. Even when only statistical information about the
eavesdropper's channel is known, substantial secrecy rate
improvement is observed. Obviously, transmit and receive jamming can be combined to further
improve the secrecy rate in a MIMO wiretap channel. In addition, both of
these jamming techniques can be directly applied to the relay jamming schemes, where a set
of trusted half-duplex or full-duplex relays can help the communication between legitimate terminals while
introducing jamming signals to the eavesdroppers. The half-duplex relays could introduce artificial noise
to the eavesdroppers while transmitting the signal to the legitimate users. On the other hand, the full-duplex relays
could transmit jamming signal, while receiving the signals from the source. Specifically, the artificial noise approach
with half-duplex relays can be considered as the transmit jamming, whereas the receive jamming represents the full-duplex relay based jamming scheme.

\section{MIMO Jamming: Game Theoretic Approach}
\begin{figure}[t]
\begin{center}
   \subfigure[]{\includegraphics[width=0.25\textwidth, angle=90]{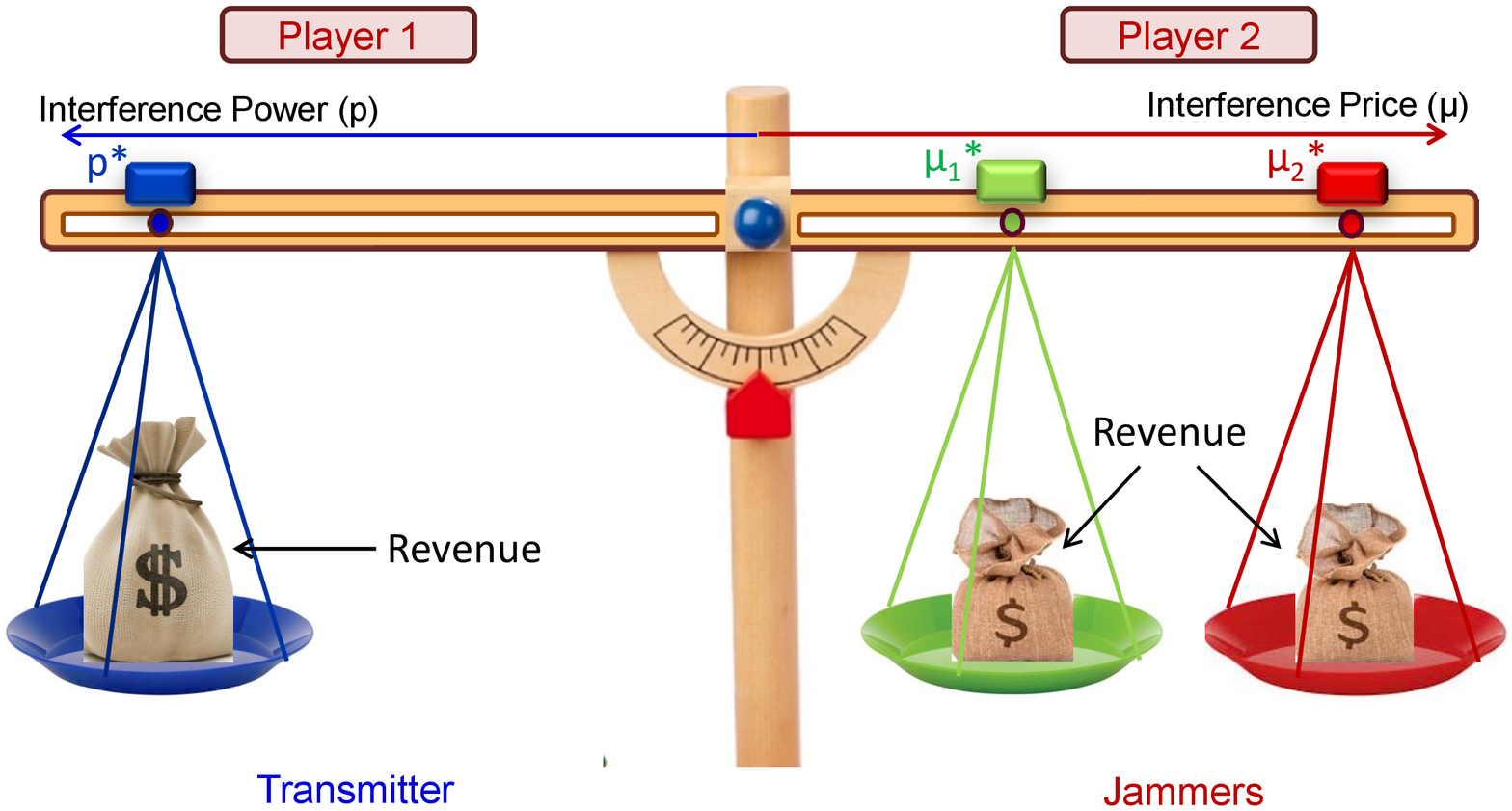}}
\end{center}
\caption{(a) A MIMO secrecy network with multiple eavesdroppers and private jammers.
(b) The concept of equilibrium for the transmitter-private jammers game.}\label{fig:Game_theory}
\end{figure}

In this section, we review game theoretic based jamming approaches for physical layer security. The centralized transmitter-receiver based jamming schemes discussed in the previous section might not be able to achieve the required performance under all circumstances due to channel conditions and the strong SI. In this scenario, external jammers can be employed to improve the quality of the secure communications by introducing jamming signals to the eavesdroppers as shown in Fig. \ref{fig:Game_theory} (a) \cite{Han_TVT_J12,Chu_TVT_J14,Cuma_JSTSP_J16}. However, these external (private) jammers charge for their dedicated jamming services from the transmitter based on the amount of interference caused at the eavesdroppers. In order to compensate these prices, the transmitter introduces charges for its enhanced secure transmission from the legitimate users. In this scenario, both the transmitter and the private jammers compete to maximize their revenues by providing higher secrecy rates at the legitimate users and selling the interference to the transmitter, respectively.\\
\indent In order to analyze the interactions between the transmitter and the private jammers, game theory provides the mathematical structure and concepts to formulate this scenario into different games, where both the transmitter and the private jammers will be the players of the game \cite{Han_Game_theory_B10}. In addition, analytical results obtained through these games  based on the strategic decisions of the players will help to conclude whether there is an equilibrium in their revenues for the proposed game as shown in Fig. \ref{fig:Game_theory}(b), at which both players achieve their maximum revenues. On the other hand, achieving these equilibria among the players might introduce more complexity in the network, even though they are beneficial for all players. To circumvent these complexity issues, game theory also facilitates to develop distributed and low complexity based implementation to attain these equilibria. Based on whether it is a collaboration or competition among the players, these games can be classified into cooperative and non-cooperative games, respectively \cite{Han_Game_theory_B10}.\\
\indent A secrecy network model is considered with one transmitter and one legitimate user, where private jammers help to improve the secure transmission by causing interference to the eavesdroppers as shown in Fig.  \ref{fig:Game_theory}(a). The transmitter, the legitimate user and the private jammers are all equipped with multiple antennas. As mentioned earlier, the interaction of buying interference from private jammers and announcing interference prices to the transmitter can be formulated into a \emph{Stackelberg} game, where the private jammers and the transmitter are the \emph{leaders} (Player 1) and the \emph{follower} (Player 2) of the game, respectively \cite{Han_Game_theory_B10}. In order to study this game, the best responses of both the players should be derived, where the best interference requirements for a given interference price and the best interference price for a given interference requirements will be obtained at both the players. Based on these best responses, the \emph{Stackelberg} equilibrium can be achieved by deriving the optimal interference requirement and the interference prices. At this equilibrium as shown in Fig. \ref{fig:Game_theory}(b), both of the players will achieve their maximum revenues and the deviation of any player from this equilibrium will cause a loss in their revenues. This equilibrium can be implemented by exchanging the associated channel responses between the transmitter and the private jammers. However, this might introduce more complexity in the network. Therefore, a distributed implementation of this equilibrium would be more appropriate to reduce the complexity, where the interference prices offered by the jammers and the interference requirements at the transmitter can be iteratively updated \cite{Han_TVT_J12}.\\

\indent The secrecy network might consist of multiple legitimate multi-antenna transmitter-receiver pairs as well as a friendly jammer in the presence of multiple eavesdroppers. In this scenario, each pair tries to maximize its secrecy rate with the help of the friendly jammer by introducing interference to the eavesdroppers. However, the jamming signals will cause interference not only to the eavesdroppers but also to the destinations, which will degrade the secrecy rate performance. Therefore, the jammer should distribute its power among the users such that the secrecy rate performance is improved. These interactions between the pairs and the jammer can be formulated into an \emph{auction} game, where the transmitters and the jammer will be the bidders and auctioneer \cite{Han_Game_theory_B10}, respectively. The transmitters will submit their bids to the jammer depending on the payment of the corresponding jammer-power and the secrecy rate improvement whereas the jammer will determine the optimal power allocations between the users based on these bids. For this game, a distributed solution can be developed by exploiting the distributed auction theoretic approach, where the bids from the transmitters and power allocations between the users will be updated iteratively.

\section{Enhancing physical layer Security in Wireless Energy Harvesting Networks}\label{sec:Enhancing PHY-layer Security in WEH-enabled Networks}
Since there has been an upsurge of research interest as well as emerging applications for radio-frequency (RF) signal-enabled wireless energy transfer, which avails the broadcasting and far-field radiative properties of electromagnetic (EM) wave to power wireless devices in particular, while transferring information, new challenges as well as opportunities for physical layer security begin arising in these WEH-enabled networks. In this section, we discuss the-state-of-art technologies enhancing physical layer security for one important class of WEH application, i.e., \emph{wireless powered communication network}  (WPCN).

\begin{figure}[t]
\centering
 \epsfig{file=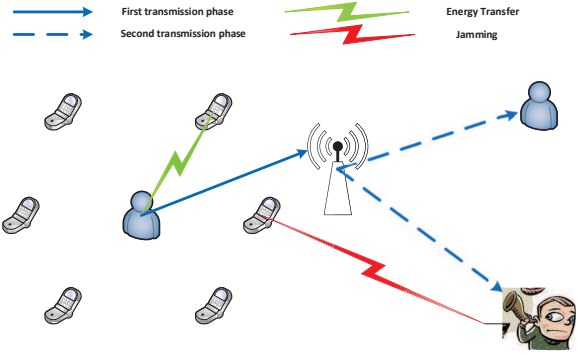, scale=0.3}
\caption{A WPCN model employing harvest-and-jam (HJ) protocol.} \label{fig:PHY-layer security for WEH}
\end{figure}

Physical layer security issues in the rapidly developed cooperative networks such as device-to-device systems, relay networks  etc., have already drawn significant attention and inter alia, CJ has been widely studied as a promising technology \cite{tekin2008general,mukherjee2014survey}. However, the benefits of CJ would be quite compromised if those energy-limited potential helpers are unwilling to cooperate. In the following, we introduce how this bottleneck could be broken with self-sustainable terminals in a WPCN. Benefiting from dense radio-frequency (RF) signals of  increasing amount of data transmission in the cooperative networks, a newly designed two-phase protocol, i.e., \emph{harvest-and-jam} (HJ) was proposed in \cite{xing2015HnJ} to achieve secrecy transmission by CJ and yet not to add extra power cost. Specifically, as illustrated in Fig.~\ref{fig:PHY-layer security for WEH}, in the first transmission phase, a single-antenna transmitter sends confidential information to a multi-antenna amplify-and-forward (AF) relay with conventional power supply and simultaneously transfers power to a group of idle multi-antenna users serving as helpers; in the second transmission phase, the AF relay amplifies, and forwards the message to the legitimate receiver under the protection of jamming, which is generated from each of the helpers by its harvested power in the previous transmission phase. The secrecy rate is maximized by optimizing the transmit beamforming matrix for the AF relay and the jamming covariance matrices for each of the helpers subject to transmit power constraints, under circumstances of perfect and imperfect CSI available at the coordinating node, respectively.
The results in \cite{xing2015HnJ} showed that the HJ scheme plays a prominent role in improving secret communications in practical scenarios of imperfect CSI. Especially, when CSIs related to the eavesdropper are hard to obtain and more imperfect than those related to the legitimate users, the optimal power allocation scheme inclines to jam the eavesdropper using all available power, and hence considerably degrades its information reception while minimizing the interference caused to the legitimate receiver. Typically, in a WPCN setup where helpers are fixed and evenly distributed around a disk centered on the transmitter with radius \(2\)m , assuming a simple channel fading model comprising Rayleigh fading and pathloss given by \((\frac{d}{d_0})^{-\alpha}\), given the reference distance \(d_0=1\)m and the attenuation factor \(\alpha=3\), if the transmit power is set to be \(0\)dBm and the energy conversion efficiency is \(50\%\), the average per-antenna harvested power at each helper is around \(62.5\mu\)W.

Besides increasing the secrecy rate (short-term metric) via adaptive power allocation over antennas for AF beamforming and jamming signal design, \cite{zhou2014WPjammer} considered a similar idea of increasing the confidential information transmission throughput (long-term metric) via a wireless-powered friendly jammer by proposing a delicately designed protocol with not necessarily equal time duration for the ``harvest'' and ``jam'' phases. Specifically, defining a ``power transfer (PT)'' block for the jammer to harvest energy from the source and an ``information transmission (IT)'' block for confidential information to be transmitted under the protection of a jamming signal, a threshold amount of energy, i.e., \(\mathcal{P}_JT\), which supports jamming using power \(\mathcal{P}_J\) for a transmission block of \(T\) time unit, is examined at the jammer's battery storage at the beginning of each transmission block. Only when this threshold is achieved, and meanwhile the source-destination (main channel) does not suffer from communications outage, the IT block with jamming starts. Otherwise, the transmission enters into either \emph{dedicated PT block } (when the threshold condition is not satisfied) or \emph{opportunistic PT block} (when communication outage over the main channel occurs). Based on this protocol design, four types of PT\text{-}IT cycles consisting of varied combination of these blocks are characterized, and as a result, the long-term behavior of this stochastic process has been analyzed with a closed-form achievable throughput. Finally, the design parameters, \(\mathcal{P}_J\) and the fixed transmission rates are optimized to further maximize the secrecy throughput under the constraint of the secrecy outage probability.

\section{Research Challenges}
\indent In this section, we provide future research challenges associated with the physical layer security jamming schemes discussed in the previous sections.
\subsection{Theoretical limits}
\indent Though existing works have proposed a variety of jamming schemes to enhance security level, the secrecy capacity has not been
found even for the simple WT-HI channel in Fig. \ref{figure1}. Future work of interest is to design more intelligent coding schemes
to achieve the secrecy capacity. For example, though the AS-based scheme in \cite{xu2014general} showed that the
channel prefixing technique is crucial  to improve the achievable secrecy rate, the effects of channel prefixing has not been fully
revealed. Moreover, this paper only considers the jamming schemes based on Gaussian random coding, which cannot
achieve a positive secure DoF at high
SNR, while recent works have shown that the non-Gaussian codes (e.g., structured codes) can achieve a positive secure
DoF at high SNR. Hence the design of non-Gaussian codes is also an interesting future research area.
\subsection{Smart MIMO jamming}
\indent The majority of literature assumes ideal Gaussian signalling and
therefore the optimal jamming also uses Gaussian signalling.
However,   Gaussian signalling  is not practical and wireless
systems often employ constant envelop signalling schemes like phase-shift-keying (PSK). In
this case, it is shown that Gaussian jamming is no longer effective, and can be removed by the multi-antenna eavesdropper using blind source separation techniques such as constant modulus algorithms \cite{new-jamming}. Therefore, there is a need to design the smart jamming signals adaptive to the specific constellation used.\\
Massive MIMO is a great enabler to achieve spectrum-efficient and energy-efficient wireless communications. However, its security implication is not well understood. Our previous work shows that massive MIMO systems, if carefully designed, are actually quite robust against both passive and active attacks \cite{Gan2015_Commun_Mag}. Further investigation is needed to understand its potential to effectively jam the eavesdroppers.

\subsection{Robust jamming games}
\indent Effective jamming requires perfect CSI, which is difficult to obtain in practice. Imperfect CSI may not only degrade the performance of the legitimate communications but also results in information leakage to the eavesdroppers. In order to deal with this issue, it is necessary to consider robust transmit jamming to achieve guaranteed outage performance of the secrecy rate. Most of the game theoretic approaches proposed for physical layer security have been assumed that the players have the perfect CSI of the eavesdroppers. In order to overcome the imperfect CSI issues associated with eavesdroppers in the existing games, robust techniques should be considered for secure communications. Therefore, the development of robust jamming games by incorporating channel uncertainties or the cases of no eavesdroppers' CSI along with the corresponding analysis of equilibria would be very challenging. These robust jamming games could be formulated into \emph{Bayesian} games, which are well known for the scenarios with incomplete information. The jamming games with imperfect eavesdroppers CSI would be one of the possible interesting future directions in game theoretic based jamming for physical layer security.

\subsection{Wireless energy harvesting}
More sophisticated transmission protocols incorporating adaptive power/rate designs are expected to be studied for further improving the long-term secrecy throughput in \cite{zhou2014WPjammer}. Besides, to further motivate the potential  WEH-enabled  helpers to assist secure communications, more practical energy and communications mechanism needs to be properly designed. For example, neighbour-users of the transmitter could be self-interested and prefer to storing the harvested energy for their own use rather than help jam. In this situation, the transmitter needs to offer some spare communications resource as incentives, such as spectrum, to increase the cooperation utility and reduce the overall system cost.
\section{Conclusions}
\indent In this article, the jamming techniques for physical layer security have been discussed with different approaches. In particular, the theoretical limits of user cooperation based jamming and the practical designs of MIMO jamming as well as the game theoretic based jamming techniques have been reviewed. In addition, the WEH based jamming has also been presented for improving energy efficiency in secure communications by exploiting the wireless harvested energy to generate jamming signals to the eavesdroppers. Finally, future research challenges of these jamming schemes for physical layer security have been briefly outlined.


\begin{thebibliography}{1}
\bibitem{MISOME}
A.~Khisti and G.~W. Wornell, ``{Secure transmission with multiple
antennas I: The  MISOME wiretap channel},'' {\em IEEE Trans. Inf. Theory}, vol.~56,  no. ~7,   pp.~3088--3104, Jul. 2010.
\bibitem{HH_Chen_WC_J11}
Y. S. Shiu, S. Y. Chang, H. C. Wu, S. C. H. Huang and H. H. Chen, ``Physical layer security in wireless networks: a tutorial," \emph{IEEE Wireless Commun.}, vol. 18, no. 2, pp. 66–74, Apr. 2011.
\bibitem{Cuma_TVT_J14}
K. Cumanan, Z. Ding, B. Sharif, G. Y. Tian and K. K. Leung,, ``Secrecy rate optimizations for a MIMO secrecy channel with a multiple-antenna eavesdropper," \emph{IEEE Trans. Veh. Technol.}, vol. 63, no. 4, pp. 1678--1690, May, 2014.
\bibitem{Qaraqe_Comm_lett_J16}
H. Lei, H. Zhang, S. Ansari, C. Gao, Y. Guo, G. Pan, and K. A. Qaraqe, ``Performance analysis of physical layer security over generalized-K fading channels using a mixture Gamma distribution," \emph{IEEE Commun. Lett.}, vol. 20, no. 2, pp. 408–411, Feb. 2016.
\bibitem{tekin2008general}
E. Tekin and A. Yener, ``The general Gaussian multiple-access and two-way wiretap channels: achievable rates and cooperative jamming," \emph{IEEE Trans. Inf. Theory}, vol. 54, no. 6, pp. 2735--2751, Jun., 2008.
\bibitem{Zheng_TSP}
G. Zheng, I. Krikidis, J. Li, A. P. Petropulu, and B. Ottersten, ``Improving physical layer secrecy using full-duplex jamming receivers'', {\em IEEE Trans.  Signal Process.}, vol. 61, no. 20, pp. 4962--4974 , Oct. 2013.
\bibitem{mukherjee2014survey}
A. Mukherjee, S. A. A. Fakoorian, J. Huang and A. L. Swindlehurst, ``Principles of physical layer security in multiuser wireless networks: A survey," \emph{IEEE Commun. Surveys Tuts.}, vol. 16, no. 3, pp. 1550--1573, Third Quater, 2014.
\bibitem{Xia_TWC_J15}
H. M. Wang, T. Zheng, and X. G. Xia, "Secure MISO wiretap channels with multiantenna passive eavesdropper: Artificial noise vs. artificial fast fading," \emph{IEEE Trans. Wireless Commun.}, vol. 14, no. 1, pp. 94–106, Jan.
2015.
\bibitem{Xia_Comm_Mag_J16}
H. M. Wang and X. G. Xia,``Enhancing wireless secrecy via cooperation: Signal design and optimization," \emph{IEEE Commun. Mag.}, vol. 53, no. 12, pp. 47–53, Dec. 2015.
\bibitem{Han_TVT_J12}
R. Zhang, L. Song, Z. Han and B. Jiao, ``Physical layer security for two-way untrusted relaying with friendly jammers," \emph{IEEE Trans. Veh. Technol.}, vol. 61, no. 8, pp. 3693--3704, Oct., 2012.
\bibitem{Chu_TVT_J14}
Z. Chu, K. Cumanan, Z. Ding, M. Johnston and S. Le Goff, ``Secrecy rate optimiztions for a MIMO secrecy channel with a cooperative jammer," \emph{IEEE Trans. Veh. Technol.}, vol. 64, no. 5, pp. 1833--1847, May, 2015.
\bibitem{Cuma_JSTSP_J16}
K. Cumanan, Z. Ding, M. Xu and H. V. Poor, ``Secrecy rate optimization for secure multicast communications," \emph{IEEE J. Sel. Topics Signal Process.}, vol. 10, no. 8, pp. 1417–1432, Dec., 2016.
\bibitem{tang2011interference}
X. Tang, R. Liu, P. Spasojevic and H. V. Poor, ``Interference assisted secret communication," \emph{IEEE Trans. Inf. Theory}, vol. 57, no. 5, pp. 3153--3167, May, 2011.
\bibitem{lai2008relay}
L. Lai, and H. El Gamal, ``The relay--eavesdropper channel: Cooperation for secrecy," \emph{IEEE Trans. Inf. Theory}, vol. 54, no. 9, pp. 4005--4019, Sept., 2008.
\bibitem{xu2014general}
P. Xu, Z. Ding, X. Dai and K. K. Leung, ``A general framework of wiretap channel with helping interference and state information," \emph{IEEE Trans. Inf. Forensics Security}, vol. 9, no. 2, pp. 182--195, Feb., 2014.
\bibitem{he2014structured}
X. He and A. Yener, ``Providing secrecy with structured codes: Two-user Gaussian channels,'' {\em IEEE Trans. Inf. Theory}, vol. 60, no. 4, pp. 2121--2138, Apr. 2014.
\bibitem{iJam}
S. Gollakota and D. Katabi, ``Physical layer wireless security made fast and channel independent,''
in \emph{Proc. IEEE Int. Conf. Comp. Commun.}, Shanghai, China, Apr. 2011, pp. 1125--1133.
\bibitem{Han_Game_theory_B10}
Z. Han, D. Niyato, W. Saad, T. Basar and A. Hj{\o}rungnes, ``Game theory in wireless and communications networks: Theory, models and applications," Cambridge, UK. Cambrdige Univ. Press, 2010.
\bibitem{xing2015HnJ}
H. Xing, K.-K. Wong, Z. Chu and A. Nallanathan, ``To harvest and jam: a paradigm of self-sustaining friendly jammers for secure AF relaying,'' \emph{IEEE
Trans. Signal Process.}, vol. 63, no. 24, pp. 6616--6631, Dec. 2015.
\bibitem{zhou2014WPjammer}
W. Liu, X. Zhou, S. Durrani and P. Popovski, ``Secure communication with a wireless-powered friendly jammer,'' \emph{IEEE
Trans. Wireless Commun.}, vol. 15, no. 1, pp. 401--415, Jan. 2016.
\bibitem{new-jamming}
O. Bakr and R. Mudumbai, ``A new jamming technique for secrecy in
multi-antenna wireless networks, ''  {\em IEEE Int. Symp. Information Theory (ISIT)},  Austin, Texas, USA,  Jun. 2010, pp.2513--2517.
\bibitem{Gan2015_Commun_Mag}
D. Kapetanovic, G. Zheng and F. Rusek,``{Physical layer seucrity for massive MIMO: An overview on passive eavesdropping and active attacks},'' \emph{IEEE Commun. Mag.}, vol. 53, no. 6, pp. 21--27, Jun. 2015.
\end{thebibliography}
\end{document}